\documentclass{PoS}
\usepackage{amsmath,amssymb,graphics}
\usepackage{grffile}
\usepackage{mcite}

\graphicspath{{}}
\usepackage{array}
\newcolumntype{L}[1]{>{\raggedright\let\newline\\\arraybackslash\hspace{0pt}}m{
#1}}
\newcolumntype{C}[1]{>{\centering\let\newline\\\arraybackslash\hspace{0pt}}m{#1}
}
\newcolumntype{R}[1]{>{\raggedleft\let\newline\\\arraybackslash\hspace{0pt}}m{#1
}}
\usepackage{graphics,epsfig}
\usepackage{epstopdf}
\fontencoding{T1}
  \fontfamily{arial}
  \fontseries{m}
  \fontshape{n}
  \fontsize{20}{18}
  \selectfont

\newcommand{\sign}[0]{\text{sign}}

\usepackage{grffile}
\title{Overlap fermions on GPUs}

\ShortTitle{Overlap fermions on GPUs}
\author{\speaker{Nigel Cundy}\\
         Lattice Gauge Theory Research Center, FPRD, and CTP, Department of Physics \&
    Astronomy,\\ Seoul National University, Seoul, 151-747, South Korea\\
        E-mail: \email{ndcundy@gmx.com}}

\author{Weonjong Lee\\
         Lattice Gauge Theory Research Center, FPRD, and CTP, Department of Physics \&
    Astronomy,\\ Seoul National University, Seoul, 151-747, South Korea}

\abstract{
We report on our efforts to implement overlap fermions on NVIDIA GPUs using CUDA, commenting on the algorithms used, implemetation details, and the performance of our code.}

\FullConference{The 33rd International Symposium on Lattice Field Theory\\
                 14 -18 July  2015\\
                 Kobe International Conference Center, Kobe, Japan}

\begin{document}
\section{Introduction}
 The overlap Dirac operator~\cite{Neuberger:1998fp,*Neuberger:1997bg,*Neuberger:1998my,*Narayanan:1994gw} is the only known practical lattice Dirac operator with exact chiral symmetry.
 It has various advantages over other discretisations, including: an exact chiral symmetry, no additive mass renormalisation, a solid definition of the topology, automatic O($a$) improvement, and good topological properties. However, it also has two major disadvantages: it requires a great deal more computational power than other actions, and the algorithms (such as the HMC algorithm used to generate gauge field ensembles~\cite{Cundy:2005pi,*Cundy:2008zc,*Cundy:2007df}) are considerably more complex. Therefore, to use overlap fermions in simulations of lattice QCD, it is necessary to have both an efficient algorithm and a good numerical implementation on the best available computer hardware.
 
 A GPU cluster provides a powerful, low (financial) cost, and energy efficient supercomputer. The GPU is a massively parallel co-processor, and because of the needs of the gaming industry, the computational power of the GPU is continually increasing while the price remains stable. GPUs also tend to be more energy-efficient than an equivalent CPU cluster. However, GPUs can be adapted to any easily parallelized computation, which makes them ideally suited for lattice QCD.  NVIDIA has been particularly supportive of this effort, and have provided a high-level programming language, CUDA, as an extension to C++. Existing lattice codes can thus be easily adapted to make use of the GPU. The GPU architecture is ideally suited for overlap fermions, which are still limited to relatively small volumes and coarse lattice spacings. The main limitation of a GPU is memory bandwidth and its physical memory limitations, meaning that on larger volumes it is necessary to set up a GPU cluster, increasing the communication costs.

 These proceedings are an initial report on an attempt to write an overlap production code for the GPU architecture. Our initial intention was to extend the QUDA library~\cite{Clark:2009wm,*Babich:2010mu} (a specialised GPU library for lattice QCD) for our purposes, but we found that the Wilson-fermion centred structure (for example odd-even preconditioning, and the difficulty of merging the QUDA Wilson operator into our other codes) made it harder to write an optimal code. We therefore ending up writing our own code entirely, though parts of it are influenced by QUDA. Our code is built on the C++ Columbia Physics System library~\cite{Jung:2014ata}, which provides the various low-level CPU routines.
 
In section \ref{sec:2} we provide details of our implementation and choices made when developing the algorithm. In section \ref{sec:3}, we show the results of various numerical tests, comparing our code against both the CPU and our initial QUDA implementation, and we conclude in section \ref{sec:4}.
 
 See~\cite{Walk:2010ut,*Walk:2012aa} for another project describing a GPU implementation of overlap fermions.

 \section{Implementation Details}\label{sec:2}
 
  { The overlap Dirac operator at mass parameter $\mu$ is defined as}
\begin{gather}
D[\mu] = \frac{1}{2}(1+\mu + (1-\mu) \gamma_5\sign(K)).
\end{gather}
 $K$ is the sign function kernel. We use the Wilson operator, $K = \gamma_5 (D_W - m_w)$, with $m_W = 1.5$ and
\begin{gather}
(D_W\psi)(x)= 4-\frac{1}{2} \sum_{\mu} \left[(1-\gamma_\mu)U_\mu(x) \psi(x+a\hat{\mu}) + (1+\gamma_\mu)U^\dagger_\mu(x-a\hat{\mu}) \psi(x-a\hat{\mu}) \right]  .
\end{gather}
 For our tests, we apply 3 steps of over-improved stout smearing~\cite{Morningstar:2003gk,*Moran:2008ra} with parameters $\rho = .1$, $\epsilon = -0.25$.
  We calculate the matrix sign function using a Chebyshev polynomial approximation and deflation of the smallest eigenvector/eigenvalues pairs, $\phi_i$ and $\lambda_i$, of $K$
 \begin{gather}
 \sign(K) = \sum_m \alpha_m T_m(K) + \sum_i \phi_i \phi_i^\dagger \left[\sign(\lambda_i) - \sum_m \alpha_m T_m(\lambda_i)\right].
 \end{gather}
  $T_m$ are the Chebyshev polynomials of the first kind, and $\alpha_m$ are the appropriate coefficients to give an approximation to the sign function. 
  We used the Chebyshev polynomial to approximate the matrix sign function because it provides a good approximation (its cost is comparable to the optimal Zolotarev rational approximation), but, needing just a short three vector recurrence, it requires less memory than the five dimensional approximation or a rational approximation. GPU applications can be limited by the amount of memory. This is particularly true for overlap fermions, where for optimal performance, we need to hold the vectors for the outer inversion, preconditioning inversion, sign function approximation, and eigenvalue deflation simultaneously in the GPU or CPU memory.

   We pre-compute a number of eigenvectors $\phi$ (the number calculated is limited by the system memory; the number used is varied according to the desired precision of the matrix sign function approximation) efficiently, and with negligible overall cost, using a polynomial preconditioned implicitly restarted Lanczos routine. These eigenvectors are stored in CPU memory. We performed the deflation on the CPU while the GPU computed the matrix sign function, using pthreads to allow the two routines to run simultaneously. This proved to be more efficient than copying all the kernel eigenvectors to the GPU to deflate on the GPU. 

 Our tests are performed on a Desktop Computer, with a quad core Intel Xeon (2.5 GHz), and two NVIDIA GK110 (GeForce GTX Titan) GPUs. The details of our computer are shown in table \ref{tab:1}.  The performance of our code is primarily limited by memory bandwidth. We tested our code on two sets of lattices, at two different (partially quenched/mixed action) masses. The parameters of our configurations are shown in table \ref{tab:2}.
 \begin{table}
\begin{center}
{
\footnotesize
\begin{tabular}{|ll|}
\hline
Memory (MiB)& 6144\\
Memory Bandwidth (Global Memory) (GB/s)&288\\
Single Precision Processing Power (GFlops)&4500\\
Double Precision Processing Power (GFlops)&1500\\
\hline
\end{tabular}
}
\end{center}
\vspace{-0.5cm}
\caption{The Memory, Memory Bandwitdth and Processing power of a NVIDIA GK110 (GeForce GTX Titan) GPU.}\label{tab:1}
\end{table}

 \begin{table}
\begin{center}
{
\footnotesize
\begin{tabular}{|llllll|}
\hline
Lattice Size& Action&Sea Mass& Valance Mass& $a^{-1}$ (GeV)&  $m_\pi a$ \\
\hline
$8^3\times 32$& Overlap& 0.01&0.05&1.7&0.662(19)\\
$8^3\times 32$& Overlap& 0.01&0.01&1.7&0.366(72)\\
$20^3\times 64$& Asqtad (MILC~\cite{Bazavov:2009bb})& $0.01+0.05$&0.01&$\sim$1.6&0.276(2)\\
$20^3\times 64$& Asqtad (MILC~\cite{Bazavov:2009bb})& $0.01+0.05$&0.0033&$\sim$1.6&0.173(2)\\
\hline
\end{tabular}
}
\end{center}
\vspace{-0.5cm}

 \caption{Details of the ensembles used for our tests. All ensembles used a L\"uscher-Weisz gauge action.}\label{tab:2}
 \end{table}
  CUDA is an extension to C++ that incorporates kernels, to be run on the GPU,  as well as routines to transfer memory from the GPU to the CPU.
 The GPU serves as a very highly threaded co-processor for the CPU. The memory on the GPU is stored in various locations. Shared memory and registers have the fastest access, then the cached texture and constant memory, followed by the global memory, which holds most of the data. There are two primary memory bottlenecks: transfer from the GPU to the CPU (very slow) and transfer from global memory to the registers (very important). The GPU/CPU bottleneck is not so significant for our code, since most calculations are performed entirely on the GPU, and we can overlap CPU/GPU communication and computation.  One important optimization of the code was to merge numerous CUDA kernels together. This saves on memory bandwidth between global memory and the registers, and reduces the kernel launch overhead. The most important gain was by merging together the Wilson matrix and polynomial algebra for the matrix sign function into a single GPU kernel. Because of these unique needs, it was difficult to integrate our code into the QUDA library. We therefore wrote our own Wilson Matrix kernel, linear algebra routines, and GPU/CPU transfer routines, though based on the implementation in QUDA. Unlike QUDA, we do not use odd/even preconditioning.
 
 We tested two different routines to invert the overlap operator. The first routine, which we label as GMRESR(eigSUMR), used a relaxed nested inversion~\cite{Arnold:2003sx} with a low accuracy eigSUMR~\cite{Arnold:2003sx,Jagels,Stathopoulos:2007zi,*Cundy:2015jza} algorithm used as a pre-conditioner for recursive GMRES~\cite{Cundy:2004pza}. This is known to perform efficiently, because the bulk of the computation only requires a low-accuracy matrix sign function, but it requires the costly pre-calculation of the overlap eigenvectors (albeit to a very low accuracy), and performs poorly on large lattice volumes. The second approach, GMRES($\alpha $MGOv), followed~\cite{Brannick:2014vda}, using the Wilson operator (with a tuned $m_W$) as a pre-conditioner for the overlap operator. The Wilson operator was inverted using an adaptive multigrid algorithm, $\alpha MG$~\cite{Frommer:2013fsa}, which is a combination of inexact deflation (we used 10 inexact deflation vectors) and the Schwartz Alternating Procedure (SAP). Optimal performance for GMRES($\alpha $MGOv) required tuning $m_W$, which we accomplished by running the inverter for a fixed number of steps, using an iterative procedure to find the $m_W$ which produced the lowest residual. $m_W$ needs to be tuned once per ensemble, since  the optimal $m_W$ did not vary much from one configuration to another, and a slightly sub-optimal $m_W$ does not significantly affect the performance. This tuning of $m_W$ is considerably faster than the computation of the overlap eigenvectors. We have not yet fully optimised our $\alpha MG$ algorithm. 
 
 These tests used a Lanczos procedure to compute the overlap eigenvectors~\cite{Stathopoulos:2007zi,*Cundy:2015jza}, however this routine performs poorly since it is not possible to significantly relax the accuracy of the matrix sign function. We also tried a Jacobi-Davidson (JD) algorithm, but found that it would not work efficiently on larger lattices. JD involves inverting the overlap operator minus a guess of an eigenvalue (projected into the subspace orthogonal to the eigenvector), and the eigenvalues were too densely packed to allow any convergence. Our current approach is to use an accelerated CG minimisation of the Ritz functional for a rational approximation of the overlap operator. This transfers computation from an eigenvalue routine to an inversion, where we use the efficient GMRES($\alpha $MGOv) algorithm. Although this has proven more efficient than the other methods, we are still not completely satisfied with this algorithm, and are still searching for alternatives. Therefore we will not report on the performance of our eigenvalue routines here.
 
 \section{Numerical Results}\label{sec:3}
 Table \ref{tab:3} gives a comparison between QUDA and our code (OC) for various routines for 1 and 2 processors. We obtain a similar performance for the Wilson operator.
 We get an improved sign function performance compared to an implementation which used the QUDA Wilson operator since we could partly merge the Wilson operator and the rest of the sign function code.
\begin{table}
\begin{center}
{
\footnotesize
 \begin{tabular}{|l|ll|ll|}
\hline
Lattice& $8^3 32$  1 GPU &  $8^3 32$  2 GPU& $8^3 32$  1 GPU&  $8^3 32$  2 GPU\\
\hline
Wilson Matrix ($10^{-3}$s)& 0.236&0.195 &0.248 &0.172
\\
Sign Function (s) & 0.2282& 0.1716&0.1448 &0.1396
\\
\hline\hline 
Lattice&$20^3 64$  1 GPU &  $20^3 64$  2 GPU& $20^3 64$  1 GPU&  $20^3 64$  2 GPU\\
\hline
Wilson Matrix ($10^{-3}$s)&7.092 &3.842 &6.759 &3.625
\\
Sign Function (s)& 16.58&9.17 & 15.31&8.71
\\
\hline
\end{tabular}
}
\end{center}
\vspace{-0.5cm}
\caption{A comparison between the QUDA and our own code's performance for the Wilson Matrix and Matrix Sign function. The middle panel refers to the results on QUDA, and the right panel with our own code.}\label{tab:3}
\end{table}

\begin{figure}
\begin{center}
\begin{tabular}{c c}
\includegraphics[width = 7.5cm,height=5cm]{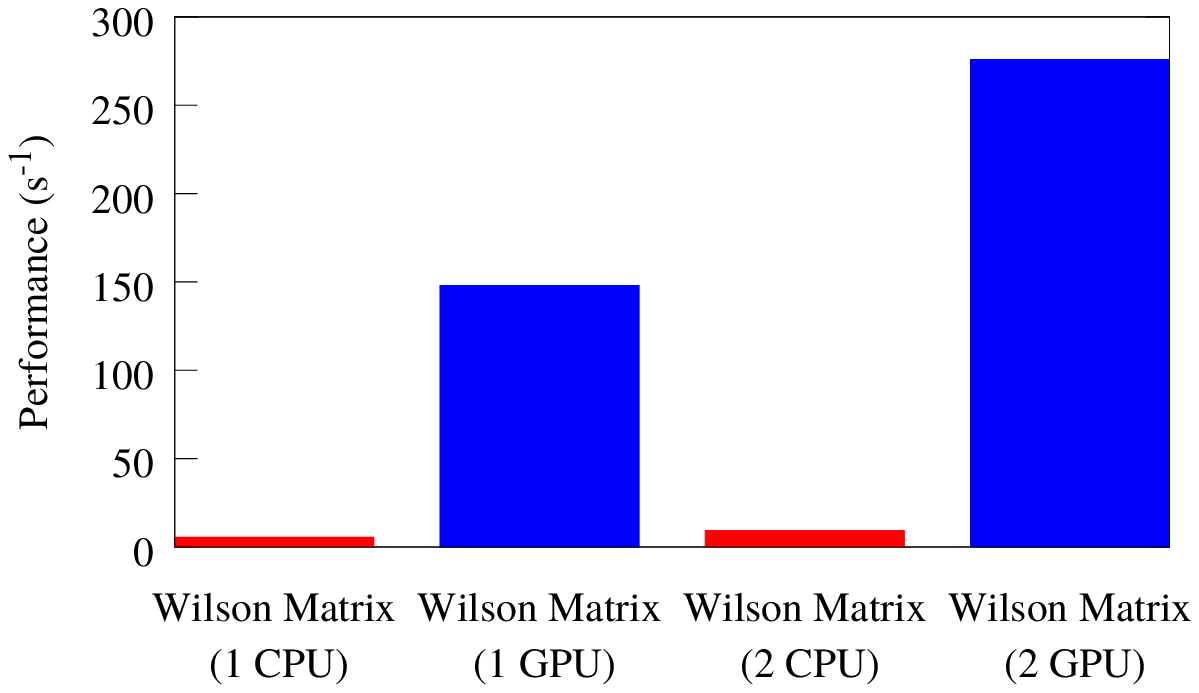}&\includegraphics[width = 7.5cm,height=5cm]{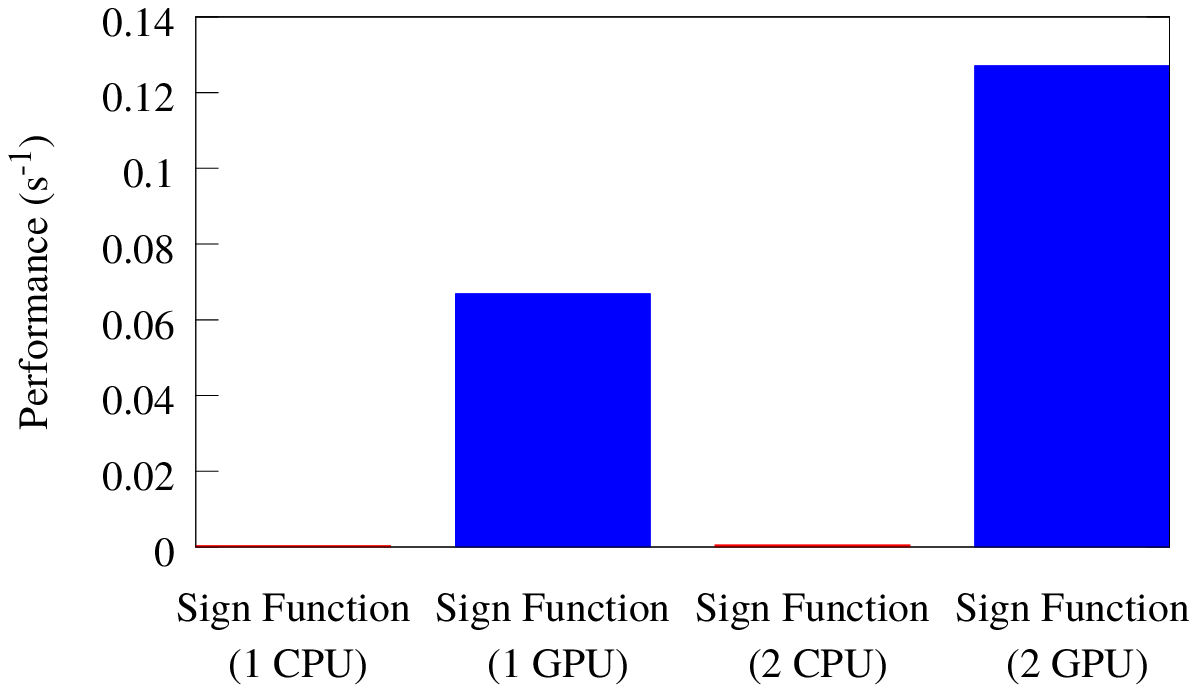}
\end{tabular}
\end{center}
\vspace{-0.5cm}
\caption{A comparison between CPU and GPU performance for the Wilson operator (left) and the Chebyshev approximation to the matrix sign function (excluding deflation) (right) on the $20^3\times 64$ lattice.}\label{fig:1}
\end{figure}

Table \ref{tab:4} gives a comparison between CPU and GPU performance for various key routines, and two of these sets of results are presented graphically in figure \ref{fig:1}. The peak performance is a comparison with the best possible time given the processing power and optimal memory bandwidth (from GPU global memory to the GPU registers). It excludes various effects which will also take up computational time, such as: the start-up time for a GPU kernel, MPI communication between processors and the GPU and CPU, and any cost caused by portions of the GPU running idle (we cannot use all the available threads for some parts of some routines because of a lack of registers on the GPU). 
\begin{table}
\begin{center}
{
\footnotesize
\begin{tabular}{|l|lll|lll|}
\hline
Routine (1 GPU) &CPU &GPU & peak& CPU  &GPU &peak
\\
\hline
Wilson Matrix ($10^{-3}s$)&5.59&0.248&43.5\%&180.4&6.759&49.9\%
\\
$\gamma_5$ ($10^{-3}s$)&0.555&0.105&19.4\%&18.92&0.885&71.8\%
\\
Chebyshev sign function(s)&8.63&0.122&42.6\%&3163&14.95&41.7\%
\\
$z = a x + b y$ ($10^{-3}s$)&1.58&0.202&15.5\%&58.2&1.50&65.0\%
\\
\hline
\hline
Routine (2 GPU) &CPU &GPU & peak& CPU  &GPU &peak
\\\hline
Wilson Matrix ($10^{-3}s$)&3.43&0.172&32.4\%&107.8&3.625&46.5\%
\\
$\gamma_5$ ($10^{-3}s$)&0.333&0.047&16.4\%&1.21&0.454&70.0\%
\\
Chebyshev sign function (s)&4.67&0.090&30.2\%&1925&7.869&40.2\%
\\
$z = a x + b y$ ($10^{-3}s$)&0.441&0.288&6.2\%&37.3&0.897&54.4\%
\\
\hline
\end{tabular}
}
\end{center}
\vspace{-0.5cm}
\caption{A comparison between GPU and CPU performance. The middle panel refers to results on the $8^3 \times 32$ lattice, and the right panel to results on the $20^3\times 64$ lattice.}\label{tab:4}
\end{table}

\begin{figure}
\begin{center}
\begin{tabular}{c c}
\includegraphics[width = 7.5cm]{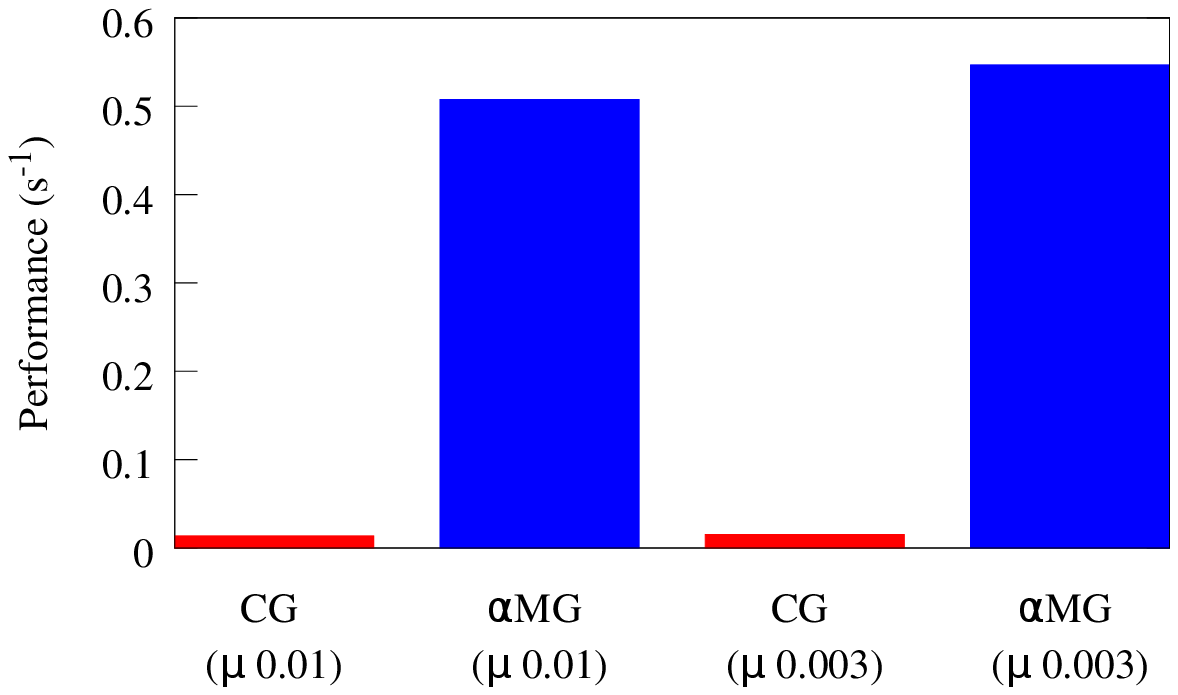}&\includegraphics[width = 7.5cm]{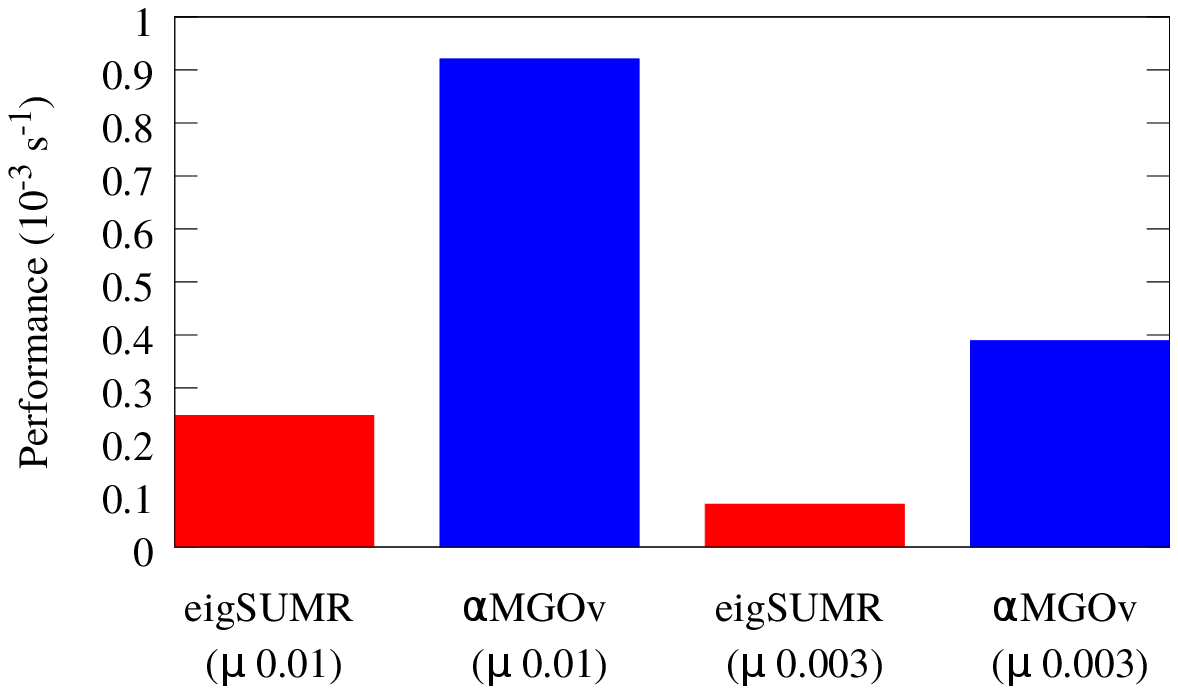}
\end{tabular}
\end{center}
\vspace{-0.5cm}
\caption{The times required for the Wilson CG and adaptive multigrid inverters (left) and the overlap inverters (right) running on 2GPUs on the $20^3\times 64$ lattice. The labels on the right plot indicate the preconditioner inside the GMRESR algorithm.}\label{fig:2}
\end{figure}

Our comparisons of the inversion algorithms are shown in table \ref{tab:5} and figure \ref{fig:2}.  We use a relative inversion accuracy of $10^{-6}$ for Wilson fermions (using single precision) and $10^{-13}$ for overlap fermions (using mixed precision). We see more than a factor of 36 gain for  $\alpha $MG over a straight-forward CG inversion for the Wilson operator on our largest volume.
 The GMRES($\alpha $MGOv) algorithm is superior to GMRES(eigSUMR) by about a factor of 5 on our largest volume and smallest quark mass. It has much better scaling with lattice volume than the eigSUMR routine.
 
 \begin{table}
 \begin{center}
{
\footnotesize
\begin{tabular}{|l l l|l l |l l|}
\hline
Volume&\#&$\mu$& CG Wilson&$\alpha MG$& GMRES(eigSUMR)&GMRES($\alpha MGOv$)
\\\hline
$8^3\times 32$ &1  &$0.05$ &0.123&0.108&9.34&19.51
\\
$8^3\times 32$ &1  &$0.01$ &0.299&0.232&10.67&45.27
\\
$8^3\times 32$ &2  &$ 0.05$ &0.114&0.091&7.09&12.93
\\
$8^3\times 32$ &2  &$ 0.01$  &0.220&0.161&9.29&29.14
\\\hline
$20^3\times 64$ &2  &$ 0.01$ &75.3&1.97&4030&1086
\\
$20^3\times 64$ &2  &$ 0.0033$ &65.76&1.83&12396&2571
\\\hline
\end{tabular}
}
\end{center}
\caption{A comparison between the performance of the  $\alpha MG$ routine against a straight CG inversion of the Wilson operator and between the overlap inversions using deflation and the Wilson operator as a pre-conditioner. The timings are given in seconds. \# refers to the number of GPUs used in the computation.}\label{tab:5}
\end{table}
\section{Conclusions}\label{sec:4}
 We have implemented a code for overlap fermions on GPUs using the CUDA programming language.
 Our code runs a factor of 200 faster than the CPU code for the matrix sign function on our larger test lattices.
 Our code scales well with the number of processors in our production environment.
 Our Wilson operator and linear algebra routines are competitive with the QUDA library.
 We have implemented inversion, eigenvalue, and conserved current routines using the latest algorithms.
 We have confirmed that the GMRES($\alpha $MGOv) algorithm performs well on large volumes on GPUs.
 We are able to perform an overlap inversion on 2 CPU/GPUs on a $20^3\times 24$ lattice at $m_\pi \sim280MeV$ in $\sim 40$ minutes.
 Eigenvalue routines for overlap fermions remain a bottleneck, but we have been working on finding a better algorithm.
 This code will eventually be used for calculations of $B_K$ and $\epsilon_K$ to check the ongoing SWME Staggered simulations~\cite{Bailey:2015tba,*Bailey:2015wta}.

 \section*{Acknowledgements}
 Computations were performed on a desktop computer at Seoul National University. 
 This research was supported by Basic Science Research Program through the National Research Foundation of Korea(NRF) funded by the Ministry of Education(2014063535). 
 W. Lee is supported by the Creative
Research Initiatives Program (No. 2015001776) of the NRF grant funded by the Korean government (MEST), and acknowledges the support from the KISTI supercomputing center through the strategic support program for the supercomputing application research (No. KSC-2014-G3-002).

  \bibliographystyle{JHEP_mcite.bst}
\bibliography{weyl}

\end{document}